\documentclass[epsf,prb,twocolumn,showpacs,nofootinbib]{revtex4-1}

\usepackage[pdftex]{graphicx}
\usepackage{dcolumn}
\usepackage{bm}
\usepackage{epsfig}
\usepackage{latexsym}
\usepackage{amsmath}
\usepackage{amsfonts}
\usepackage{amssymb}
\usepackage{color}
\usepackage{array}
\usepackage{framed}

\begin{document}

\title{Emergent gravity of fractons:\\
Mach's principle revisited}
\author{Michael Pretko\\
\emph{Department of Physics, Massachusetts Institute of Technology,
Cambridge, MA 02139, USA}}
\date{May 27, 2017}

\begin{abstract}
Recent work has established the existence of stable quantum phases of matter described by symmetric tensor gauge fields, which naturally couple to particles of restricted mobility, such as fractons.  We focus on a minimal toy model of a rank 2 tensor gauge field, consisting of fractons coupled to an emergent graviton (massless spin-2 excitation).  We show how to reconcile the immobility of fractons with the expected gravitational behavior of the model.  First, we reformulate the fracton phenomenon in terms of an emergent center of mass quantum number, and we show how an effective attraction arises from the principles of locality and conservation of center of mass.  This interaction between fractons is always attractive and can be recast in geometric language, with a geodesic-like formulation, thereby satisfying the expected properties of a gravitational force.  This force will generically be short-ranged, but we discuss how the power-law behavior of Newtonian gravity can arise under certain conditions.  We then show that, while an isolated fracton is immobile, fractons are endowed with finite inertia by the presence of a large-scale distribution of other fractons, in a concrete manifestation of Mach's principle.  Our formalism provides suggestive hints that matter plays a fundamental role, not only in perturbing, but in creating the background space in which it propagates.
\end{abstract}
\maketitle

\tableofcontents

\section{Introduction}

Much attention has been dedicated in recent years to the study of highly entangled phases of matter, such as spin liquids\cite{lucile} and fractional quantum hall systems\cite{fqh,fqh2}, which are characterized not by symmetry properties, but rather by the pattern of long-range quantum entanglement in their ground state.  (In the gapped case, long-range entanglement is usually referred to as ``topological order".\cite{topo})  One of the unifying features of such highly entangled phases of matter is their description in terms of gauge theories.  These emergent gauge theories can arise in numerous different ways: as a means of imposing spin-ice rules, via flux attachment procedures, or via parton constructions, among others.  No matter how they arise, gauge theories play a central role in our understanding of highly entangled phases.

Until recently, attention was mostly focused on the familiar vector gauge fields, since little seemed to be gained by going to higher rank tensors.  In three or fewer spatial dimensions, the ``higher form" gauge theories (antisymmetric tensors) are either unstable to confinement or provide dual formulations of vector gauge theories, and do not correspond to fundamentally new phases of matter.\cite{analytic,numerical,soojong,gerbe}  The other case one might consider is that of symmetric tensor gauge fields, which fall into the category of ``higher spin" gauge theories, since they have massless excitations of spin 2 or higher.  Historically, such theories have been notoriously difficult to formulate beyond the free field level.  For many years, all consistent interacting theories basically fell into two classes: gravity (the spin-2 case) and variants of Vasiliev theory, which involves an infinite tower of all possible higher spin fields.\cite{vasiliev}  However, recent work has shown that it is actually possible to consistently formulate an interacting theory of any higher spin gauge field in $(3+1)$ dimensions, but that it requires the existence of particles restricted to motion along certain lower-dimensional subspaces.\cite{sub,genem,alex}  This same phenomenon of restricted mobility had earlier found concrete realization in a condensed matter setting, first in work due to Chamon\cite{chamon} and in several other models\cite{bravyi,cast,yoshida}, including Haah's code.\cite{haah,haah2}  These ideas were later more systematically developed in the Vijay-Haah-Fu models.\cite{fracton1,fracton2}  As a limiting case, some subdimensional particles are restricted to a 0-dimensional subspace and are totally immobile.  These unconventional particles were given the name ``fractons"\cite{foot1,fracton1}, a topic which has been particularly active recently.\cite{williamson,han,sagar,prem,hsieh}  Fractons find a natural setting in the language of higher spin gauge fields, with the ``generalized lattice gauge theory" introduced in Reference \onlinecite{fracton2} serving as the discrete lattice analogue.  In a sense, the dual mysteries of fractons and higher spin gauge fields are two sides of the same coin.

For spin higher than 2, it seems plausible that we need some new physical ingredient, such as fractons, considering the apparent elusiveness of consistent theories.  But it seems somewhat surprising that such exotic particles should arise even in a theory of rank 2 tensors, where we have a (comparatively) much better understanding, in the form of gravity.  There are some differences in details between the rank 2 theories discussed in the condensed matter literature and conventional Einstein gravity, but they are insufficient to explain the seemingly drastic difference in properties between fracton theories and gravitational theories.  In fact, previously proposed emergent gravity models \cite{gu,gu2,cenke1,cenke2,horava} were secretly the first fracton models, though this escaped notice at the time they were first studied.  Importantly, we will discuss how the key features which lead to fractonic behavior also exist in a gravitational theory, so we must set out to reconcile our understanding of these two phenomena.  We will here do some of the basic work towards that goal by examining the simplest possible rank 2 fracton model in $(3+1)$ dimensions and elucidating how its properties can be understood in terms of emergent gravity.  A concrete lattice model will be referenced, though the main conclusions will not be lattice dependent.

The key feature which is shared between fracton models and gravity is the existence of extra conservation law(s), beyond the more familiar ones pertaining to energy, charge, and linear and angular momenta.  In a fracton model, there are additional ``higher moment" conservation laws, which cause interplay between charge and position.\cite{sub}  When the conservation laws become strict enough, particle mobility becomes entirely disallowed, and the excitations become fractons.  As we discuss in more detail below, the direct analogue in a gravitational theory is the conservation of center of mass in the system's rest frame, which in an appropriate sense is the Noether charge associated with boost symmetry.  This conservation law seems almost trivial, and it is often taken for granted since, at the classical level, it follows immediately from conservation of momentum.  But as we will discuss, in a quantum system with a preferred center of mass, single-particle momenta will no longer be good quantum numbers at all.  Furthermore, the conservation of center of mass is a \emph{local} conservation law, not simply a global one, which will have important consequences.  In the toy models we will consider here, center of mass conservation becomes a law with a life of its own.  In fact, we will discuss how this conservation law and the principle of locality lead immediately to a gravitational attraction.

In order for a particle to move, it must exchange center of mass information with the rest of the system, so as to conserve the total center of mass.  In the fracton model we will consider, we can see this happen directly.  An isolated fracton is completely immobile.  When there are multiple fractons present, however, some limited fracton mobility can occur through the virtual exchange of the emergent ``center of mass" quantum number between two fractons.  Two fractons can effectively push off of each other by exchanging a mobile particle carrying the center of mass.  The amplitude for such two-body mobility processes will be proportional to the propagation amplitude of the virtual particle between them, which decays to zero at large separation.  The propagation speed of fractons therefore drops as the separation increases, amounting to an effective attraction between them, which we will show plays the role of a gravitational force.  At large separation, the propagator of the virtual particles between the fractons approaches zero, indicating that the particles can no longer exchange any significant amount of center of mass information.  At this point, the hopping matrix elements go to zero and the particles become immobile, recovering the physics of fractons.

So if an isolated particle has no mobility, how does a normal gravitational particle move around our universe so well?  The important clue is that this particle is not truly isolated, but rather exists in a universe filled with a large-scale distribution of other gravitational sources, such as baryonic particles, dark matter, and dark energy (which we will collectively refer to as ``matter," for simplicity).  This distribution of matter effectively acts as a bath for the exchange of center of mass.  We will see in our toy model that a particle's hopping matrix elements, and therefore its inertial mass, are directly determined by the distribution of other matter present.  We therefore have an explicit manifestation of Mach's principle: the inertia of a particle is not intrinsic to the body itself, but rather is given to the particle by its interaction with the rest of its emergent universe.\cite{mach,mach2}  In this sense, the immobility of isolated fractons in a rank 2 gauge theory is a direct consequence of Mach's principle: fractons cannot move because they do not have any ``universe" to move against.

We will analyze the details of the two-body problem to see the effective attraction explicitly, showing that it matches up with the expected properties of the gravitational force, such as being always attractive and having a geometric interpretation in terms of a quasi-geodesic principle.  We will find that the gravitational force is generically short-ranged in fracton models like the toy model considered here, but the power-law behavior of Newtonian gravity can be recovered under certain conditions.  For example, gravity automatically becomes long-ranged upon introducing nonlinearity ($i.e.$ letting the graviton carry the gravitational charge).

Our formulation of the geodesic principle will also lead us to the interesting conclusion that matter is responsible not only for creating perturbations to some existing flat background metric, but also in generating the background metric itself.  When there are no particles present, the emergent space of the model is pulled apart into a set of independent isolated points.  It is only in the presence of matter that these points coalesce into a smooth geometry.

For the uncomfortable high energy reader, we also include an appendix discussing the ways in which the present work circumvents the Weinberg-Witten no-go theorem on this type of emergent gravity model.

\section{The Model}

\subsection{The Toy Model}

We will consider here a model which was studied in detail in earlier work, where it played the role of the simplest gapless fracton model.\cite{sub,genem,alex}  We will keep the present treatment self-contained, but we refer the reader to the older sources for a more in-depth analysis of certain details.  In this work, we will show that the model serves as a simplified toy version of emergent gravity.  We explain the fractonic behavior first, discussing the relationship with Einstein gravity in the next section.  We focus on the lattice-independent concepts in the main text, while the specific lattice implementation is described in Appendix A.

We take our fundamental degrees of freedom to be those of a rank 2 symmetric tensor field $A_{ij}(x)$ existing throughout a three-dimensional space.  (All indices are spatial.)  This tensor has a canonical conjugate variable $E_{ij}(x)$ which is also a symmetric tensor.  The symbols $A$ and $E$ are chosen intentionally to make connection with the vector potential and electric field of electromagnetism, which are canonical conjugates.  As emphasized in previous work, the primary factor controlling the properties of a gauge theory is the generalized Gauss's law, which in turn defines the gauge symmetry and thereby determines the rest of the theory.  For the model we will consider here, the defining Gauss's law takes the form:
\begin{equation}
\partial_i\partial_j E^{ij} = \rho
\label{gauss}
\end{equation}
where repeated indices are summed over.  (The use of ``upper" and ``lower" indices is purely for bookkeeping purposes.  Raising and lowering is done with the flat metric $\delta_{ij}$.)  The quantity $\rho$ represents an emergent scalar charge.  We will see in the next section that this equation is a proxy for the $00$ component of the linearized Einstein equations, and $\rho$ is analogous to the emergent energy, $T_{00}$.  For simplicity, we allow our gauge field $A_{ij}$ to be compact, so that this charge is quantized.  (Some details regarding compactness can be found in Appendix A.)  Equation \ref{gauss} can actually be regarded as the \emph{definition} of charge in this model, in the same way that Gauss's law can be regarded as the definition of charge in a conventional $U(1)$ Maxwell theory.\cite{u1}  In the ground state, which is free of charges, we have the constraint $\partial_i\partial_j E^{ij} = 0$.  In turn, this constraint implies gauge invariance under the transformation:
\begin{equation}
A_{ij} \rightarrow A_{ij} + \partial_i\partial_j \alpha
\end{equation}
for gauge parameter $\alpha$ with arbitrary spatial dependence.  This gauge invariance protects the gaplessness of propagating gauge degrees of freedom, which include an emergent graviton.\cite{foot2}  Keeping only the most relevant gauge-invariant terms, the Hamiltonian governing the gapless gauge modes takes the quasi-electromagnetic form:
\begin{equation}
H = \frac{1}{2}\int(E^{ij}E_{ij} + B^{ij}B_{ij})
\end{equation}
where the integral is over three-dimensional space, and $B_{ij} = \epsilon_{iab}\partial^a A^b_{\,\,\,j}$ is an appropriately defined gauge-invariant ``magnetic" tensor, which leads to a linear graviton dispersion.  From this Hamiltonian, we can derive generalized tensor Maxwell equations which will serve as the toy Einstein equations for this model.  But as we will discuss, almost all of the important physics of gravitation follows directly from Equation \ref{gauss}, which is the analogue of the $T_{00}$ component of the Einstein equations.  The full set of  Einstein-Maxwell equations can be found in Appendix A.  The precise details of the Hamiltonian and the corresponding gauge field dynamics will not be of much concern to us here.  We refer the reader to Appendix A for more precise lattice details and to Reference \onlinecite{genem} for a more detailed account of the generalized electromagnetism of this phase.

While the gapless gauge sector obeys $\partial_i\partial_j E^{ij} = 0$, we also have excited states violating this constraint, corresponding to the presence of charges, via Equation \ref{gauss}.  This generalized Gauss's law immediately allows us to identify the charge conservation laws of the system.  For example, we have a conservation of total charge, just as in ordinary electromagnetism.  The total charge can be written as:
\begin{equation}
\int \rho \,\,= \int \partial_i\partial_j E^{ij} = (\mathrm{boundary\,term})
\end{equation}
where the integrals are over the three dimensions of space.  We are integrating a total derivative, which gives us a boundary term.  In particular, if we focus on a closed manifold, the total charge must be zero.  Either way, the physical conclusion is the same: charge is always conserved in the bulk.  No local operator in the bulk of the system can change the total charge.  Any operator that creates charged particles must create an equal number of positive and negative charges.  The total charge of the system can only change if charges flow in or out through the boundary.

Charge conservation should be fairly familiar.  But we can use similar logic to identify a sneakier conservation law in this system.  Consider the total dipole moment of charges in the system:
\begin{align}
\begin{split}
\int x^i\rho &= \int x^i\partial_j\partial_k E^{jk} = \\&-\int \partial_k E^{ik} + (\mathrm{b.t.}) = (\mathrm{b.t.})
\end{split}
\end{align}
where $x^i$ is the spatial coordinate and ``(b.t.)" denotes a boundary term.  We integrated by parts in the middle step, then integrated the total derivative.  At the end of the day, we once again end up with only a boundary term.  We see that the total dipole moment also has a conservation law, changing only when charges enter or leave the system through the boundary.  In addition to respecting charge conservation, any local operator in the bulk must also respect dipole moment conservation.

\begin{figure}[b!]
 \centering
 \includegraphics[scale=0.25]{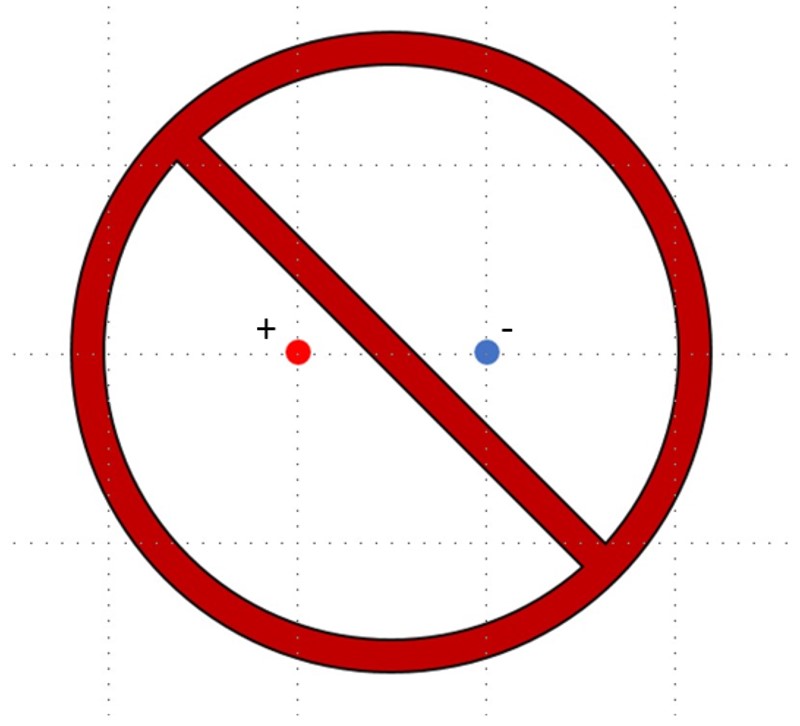}
 \caption{A dipole (particle hopping) operator is disallowed by the dipole conservation law.}
 \label{fig:dipole}
 \end{figure}

\begin{figure}[t!]
 \centering
 \includegraphics[scale=0.3]{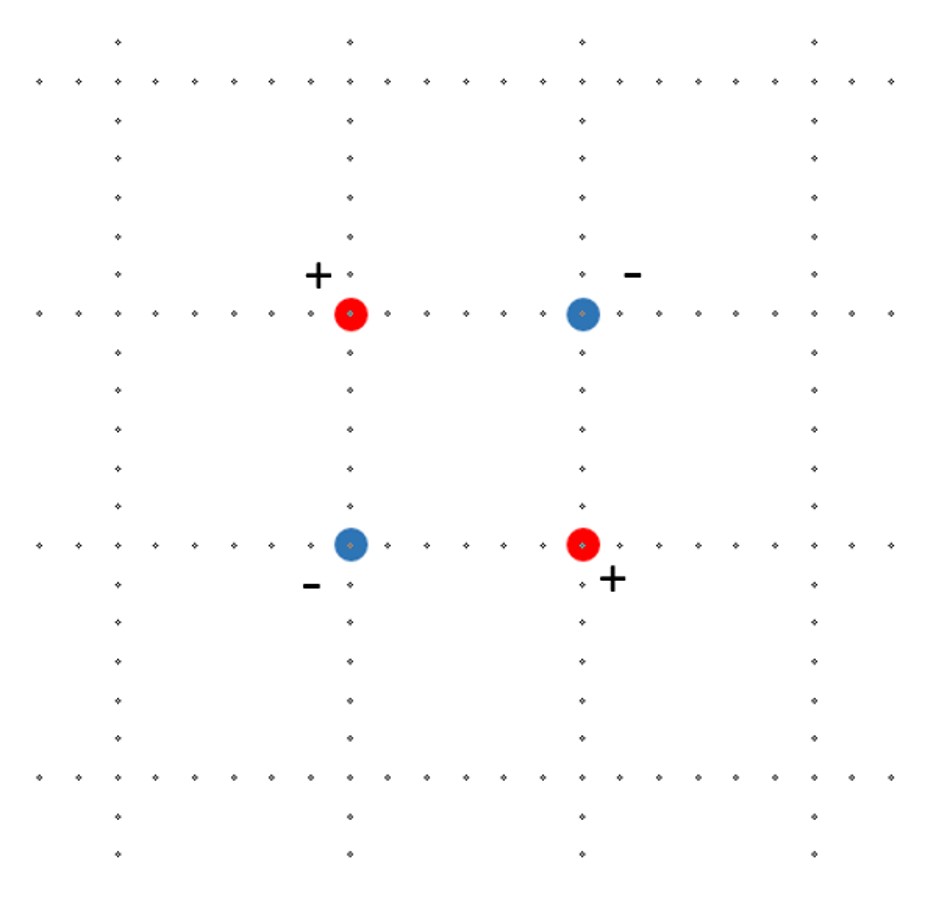}
 \caption{One acceptable creation operator is a square quadrupole configuration, amounting to a transverse dipole hop.}
 \label{fig:quadrupole1}
\end{figure}

\begin{figure}[t!] 
 \centering
 \includegraphics[scale=0.35]{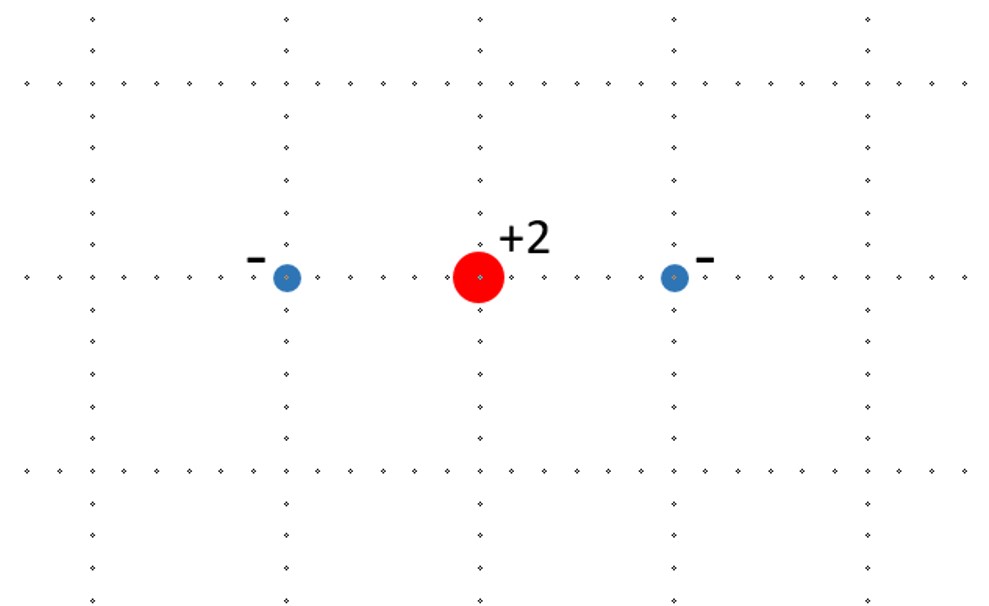}
 \caption{Another acceptable operator is the linear quadrupole configuration, amounting to a longitudinal dipole hop.}
 \label{fig:quadrupole2}
 \end{figure}

This conservation law immediately puts some severe restrictions on the sorts of local operators which can occur.  No operator can create particles in a dipole configuration, such as in Figure \ref{fig:dipole}, since this is in violation of the conservation of the global dipole moment.  Since such dipole creation operators also serve as particle hopping operators (lowering the charge at one site and increasing it at the next), we see that the conservation laws rule out traditional particle mobility, making these fractonic excitations.  The only allowable particle creation operators are those which create particles in purely quadrupolar configurations, such as those seen in Figures \ref{fig:quadrupole1} and \ref{fig:quadrupole2}.\cite{foot3}  Such operators can ``hop" a particle, but only at the expense of creating additional particles, which is energetically unfavorable.  A fracton would need to constantly absorb or emit other particles at each step of its motion to continue hopping, as opposed to normal particles which can propagate freely all by themselves.

While a charged particle is fractonic, neutral bound states of such fractons can actually be mobile.  In particular, consider a dipolar bound state, consisting of a positive charge and a negative charge.  Note that, despite its neutrality, this is not a trivial excitation, as it cannot be locally created from the vacuum due to the dipole conservation law.  Due to its neutrality, such a bound state is free to move around the system, though it is forced to maintain a fixed dipole moment.  In fact, the quadrupolar operators of Figures \ref{fig:quadrupole1} and \ref{fig:quadrupole2} are exactly the hopping matrix elements for such mobile dipoles.  The square quadrupole of Figure \ref{fig:quadrupole1} corresponds to a transverse hopping operator, while the linear quadrupole of Figure \ref{fig:quadrupole2} corresponds to the longitudinal hopping operator.

\subsection{Comparison with Einstein Gravity}

The preceding model is certainly interesting, but what does it have to do with gravity?  Well, one of the original motivations for writing down such tensor gauge theories was the fact that gravity is also described by a symmetric tensor, in the form of the metric $g_{\mu\nu}$.  For simplicity, let us only consider linearized Einstein gravity, where we write the metric in the form $g_{\mu\nu} = \eta_{\mu\nu} + h_{\mu\nu}$, with $\eta_{\mu\nu}$ being the Minkowski metric and $h_{\mu\nu}$ being a small perturbation.  The action for linearized Einstein gravity is invariant under the gauge transformation $h_{\mu\nu} \rightarrow h_{\mu\nu} + \partial_\mu \xi_\nu + \partial_\nu \xi_\mu$, which is a descendant of diffeomorphism invariance.\cite{foot4}  Components with a time index, $h_{00}$ and $h_{0i}$, have non-dynamical equations of motion, acting as Lagrange multipliers to enforce gauge constraints, in the same way that $A_0$ acts as a Lagrange multiplier enforcing Gauss's law in Maxwell theory.\cite{gr,foot5}  The remaining physical degrees of freedom are those of the spatial symmetric tensor $h_{ij}$.  We call its canonical conjugate $\pi_{ij}$.  For linearized Einstein gravity, one can show\cite{gr} that the gauge constraints on these variables are given by:
\begin{equation}
\partial_i\pi^{ij} = T^{0j}
\label{gravmom}
\end{equation}
\begin{equation}
\partial_i\partial_j h^{ij} - \partial^2 h^i_{\,\,\,i} = T^{00}
\label{gravgauss}
\end{equation}
where $T^{\mu\nu}$ is the stress-energy tensor.  These are essentially just the $0j$ and $00$ components of the Einstein equations.  (We have absorbed numerical constants and the gravitational coupling into $T^{\mu\nu}$ for simplicity.)  The corresponding gauge transformations are:
\begin{equation}
h_{ij}\rightarrow h_{ij} + \partial_i\xi_j + \partial_j\xi_i
\end{equation}
\begin{equation}
\pi_{ij}\rightarrow\pi_{ij}+\partial_i\partial_j\alpha - \delta_{ij}\partial^2\alpha
\end{equation}
for parameters $\xi_i$ and $\alpha$ with arbitrary spatial dependence.  Note that there are constraints and gauge transformations on both the field $h_{ij}$ and its conjugate $\pi_{ij}$, so the choice of calling $h_{ij}$ the ``gauge field" is somewhat arbitrary.  Equation \ref{gravmom} corresponds to a spatial vector charge representing an emergent momentum, while Equation \ref{gravgauss} has a spatial scalar charge representing the emergent energy.  It is this second equation which is analogous to the Gauss's law of the toy model, $\partial_i\partial_j E^{ij} = \rho$.  We have opted for a toy model with a simplified Gauss's law, as opposed to working directly with a model with the full structure of Einstein gravity, so as to isolate the essential ingredient required for gravitation, which is a conservation law which is shared between the two theories.  The equation $\partial_i\partial_j E^{ij} = \rho$ will turn out to be the minimal gauge constraint with sufficient structure to yield gravitational behavior.

The presence of two terms on the left side of Equation \ref{gravgauss} makes it superficially look more complicated than our toy Gauss's law, but having two derivatives allows us to immediately prove the exact same sort of conservation law:
\begin{equation}
\int x^i T^{00} = (\mathrm{boundary\,term})
\end{equation}
corresponding to conservation of center of mass in the rest frame.  In a sense, this corresponds to the conserved Noether charge of boost symmetry.  The generator of a boost in the $i$ direction takes the form $\int (x^iT^{00} - tT^{0i})$, which depends explicitly on time, unlike more familiar symmetry generators.  Nevertheless, we can still find a conservation law.  Energy eigenstates can be chosen to also be eigenstates of boosts, so we have:
\begin{equation}
\int x^i T^{00} = t\int T^{0i} + (\textrm{constant})
\end{equation}
on our eigenstates.  This simply implies that the center of mass\cite{foot6} is moving at a constant velocity, determined by the total momentum of the system.  However, we are free to pick the reference frame in which the total momentum of the system is zero, and the center of mass is at rest.  Since the total momentum of the system is conserved, the center of mass will stay stationary in this frame at all times, or in other words, it obeys a conservation law.

We now have an alternative interpretation of the dipole conservation law, $\int x^{i}\rho$ = constant, in our toy model.  If we simply regard $\rho$ as a ``mass" density (which is positive for particles and \emph{negative} for antiparticles), we can rephrase dipole moment conservation as ``center of mass" conservation of the emergent mass.  Since this conservation law led directly to fractonic behavior, we should somehow be able to reconcile our understanding of fractons with the consequences of center of mass conservation in a gravitational theory.  This will be the primary task of this paper.

Of course, our toy model is not a full match for Einstein gravity, which has a more complicated Gauss's law, with both scalar and vector charges.  Our toy model only has an emergent ``energy," whereas a full gravitational theory would have both emergent energy and momentum, leading to an emergent Lorentz invariance.  (It seems that such an emergent momentum will fit naturally into the model of Gu and Wen\cite{gu,gu2}, though the details will need to be carefully studied.)  Also, the details of the low-energy Hamiltonian for the gauge mode in our toy model are different from Einstein gravity.  The graviton dispersion in the toy model is still linear, but we have extra gapless gauge modes with no analogue in Einstein gravity.  Nevertheless, this difference is confined to the gauge mode sector and does not affect fractonic behavior, which is our main concern in the present work.  We focus here on the dynamics of the fractons, which appear to be insensitive to such differences.  We will find that our simplified toy model is sufficient to capture the physics of gravitational attraction.

\section{Mobility and Gravitation}

\subsection{Mobility via Center of Mass Exchange}

Now that we understand our higher moment conservation law as center of mass conservation, one might be tempted to think that fractons aren't so fractonic after all.  Given two fractons of equal charge, why not just consider a state where they move in opposite directions at the same speed?  Such a state is certainly consistent with the center of mass conservation law.  However, it will turn out to not be an eigenstate of the Hamiltonian.  Two fractons of equal charge have a definite center of mass, and the effective Hamiltonian for a fracton will depend on its distance from this center of mass.  While we still possess overall translation invariance of the system, we do not have translation invariance for an individual particle, and single-particle momentum is no longer a good quantum number.  Instead, we must work directly in position space.

\begin{figure}[b!]
 \centering
 \includegraphics[scale=0.3]{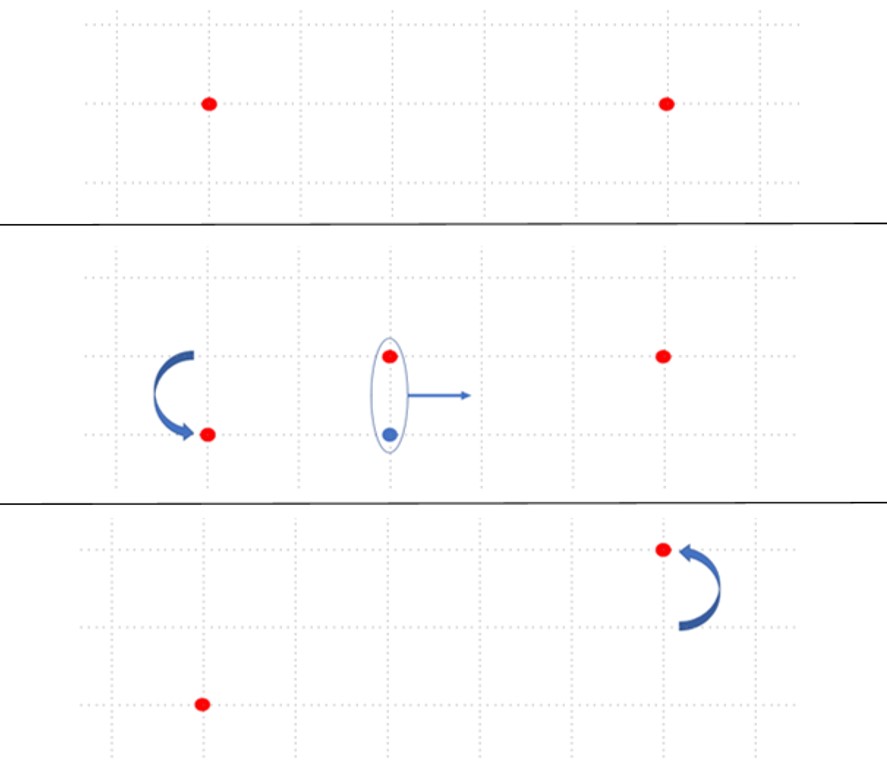}
 \caption{The left fracton can hop down by emitting a virtual dipole, which then propagates to the right fracton, where its absorption results in an upwards hop.}
 \label{fig:2part}
 \end{figure}

To begin, we examine a system with two particles of equal charge, as shown in Figure \ref{fig:2part}.  As illustrated, one of the fractons can move, but only at the expense of creating a dipole (a particle-antiparticle pair) which is mobile and can propagate to the other fracton.  If the second fracton absorbs this dipole, it will hop an equal distance in the opposite direction as the first fracton.  In this way, the fractons can move while leaving the center of mass of the system unchanged.  It is important to note that this all actually occurs as a virtual process.  The system does not have the extra energy or momentum required to create a real on-shell dipole to exchange between the fractons.  Nevertheless, such processes can (and do) occur at the virtual level.  The amplitude for such a hop, call it $t$, will depend on several things, such as the amplitudes for emission and absorption of the dipole.  But most notably, $t$ will be proportional to the propagator of the mobile dipole between the two fractons.  Suppose we write $d^\dagger(r)$ and $d(r)$ for the creation and annihilation operators of a dipole at location $r$, and let $r_1$ and $r_2$ be the locations of the two fractons.  The spatial dependence of the hopping amplitude $t$ is given by:
\begin{equation}
t(r) = \alpha \langle 0|d(r_2)d^\dagger(r_1)|0\rangle \equiv \alpha D(r)
\end{equation}
where we have defined $D$ as a shorthand for the real-space equal-time propagator, $r$ is the distance between the two fractons,\cite{foot7} and $\alpha$ is some constant.  (By the overall translation invariance of the system, the propagator only depends on $r = r_2-r_1$.)  Note that, while the two separated fractons must move precisely in unison, there are no violations of causality.  The nonzero nature of the propagator at well-separated points is a familiar issue in field theory.  Correlation of well-separated events does not imply causation.

The dipole propagator, and therefore the hopping matrix elements, will approach zero at large distances.  For the case where the dipoles have some finite energy gap $M$, hopping elements behave as $t(r)\propto e^{-Mr}$.  When the particles are separated by more than a few times $M^{-1}$, the hopping matrix elements are essentially zero, and fractonic behavior is very quickly recovered.  Or, when we have gapless dipoles, $t(r)$ falls off as a power law, which is an important case that we will discuss later.  Either way, the resulting hopping elements $t(r)$ decay as the fractons become separated.

The decay of hopping elements means that the velocity of the fractons gets smaller as they move apart, and equivalently gets larger as they move together.  This is the same effect as an attractive force between the two particles.  In fact, we will see later how this attraction has a geometric character, as expected for a gravitational force.  But note that the normal logic of attraction is somewhat turned on its head.  We normally think of the motion of the particle as intrinsic to the body, and other particles simply influence this motion by means of various forces.  In the present case, motion is inherently a cooperative effect, driven by the interaction between particles.  As the fractons approach each other, they are not speeding up through the exertion of a ``force" per se, but rather are jointly increasing each other's mobility through a more rapid virtual exchange of center of mass information.  One cannot start from momentum eigenstates and use perturbation theory to analyze changes in motion.  Rather, one must start with position eigenstates and perturb in the motion itself.

It is important to note that, while like-charged fractons experience an attractive force, oppositely charged fractons do \emph{not} experience a repulsion.  If the two fractons had opposite charges, then ``center of mass" conservation would imply that the fractons must move in the \emph{same} direction upon exchanging a dipole, not opposite.  This corresponds to the fact that two opposite charges must always maintain a fixed distance from each other to satisfy the conservation law.  As such, oppositely charged fractons can never attract or repel, and the force between them is exactly zero.

The gravitational attraction of like charged fractons is an inevitable consequence of the local nature of the center of mass quantum number.  Whenever a particle moves, it is changing its own contribution to the total center of mass of the system.  Since the total center of mass must remain exactly conserved at all times, the rest of the universe must respond in unison by moving in the opposite direction.  When there are other particles nearby, the exchange of center of mass information can occur rapidly.  But when the fracton is well-separated from other matter, the locality of the interactions prevents the efficient exchange of center of mass information between the involved parties, leading to a slowdown of the motion.

Note that the standard formulation of a non-gravitational theory with center of mass conservation is a mildly special case, in that center of mass is only conserved globally.  For example, in a free field theory, the eigenstates are described in the momentum basis, not position.  Such a state is not in a center of mass eigenstate, but rather is a superposition of all possible locations of the center of mass.  The system does not have a definite center of mass for particles to be attracted towards, and there is on average no gravitational force.  Nevertheless, any realistic collection of particles created in the lab has a definite center of mass, breaking the single-particle translation invariance.  In any theory where the locality of center of mass is respected, we expect such particles to gravitate towards their center of mass.

The disconnect between this perspective on gravity and the familiar momentum eigenstates of free field theory is in some ways reminiscent of the theory of superconductivity, where one has to choose between working with eigenstates of particle number or eigenstates of phase.  The particle number basis is more familiar, but it tends to mask the important physics, since the superconducting order parameter vanishes in any particle number eigenstate.  Only in the phase basis is it clear that superconductivity is a fairly ubiquitous phenomenon at low temperatures.  We face a similar choice between momentum and position eigenstates in a quantum gravity theory.  The momentum basis tends to be more familiar, building on intuition from free field theory, but the position basis leads to much clearer intuition on the origin of the gravitational attraction.

\subsection{Mach's Principle}

We have argued above that the hopping matrix elements of a particle fall off to zero as a particle becomes separated from all other particles in the system.  In other words, its effective mass becomes infinite.  This seems to be at odds with our normal understanding of gravitational particles.  After all, we see particles moving around all the time in our everyday life, with a seemingly finite inertia, $i.e.$ finite hopping matrix elements.  How can we obtain such physics in our toy model?  In order for a particle to move at constant speed, it must have some bath with which it can constantly exchange center of mass information.  Let us examine some possible sources of such a center of mass bath.

Naively, one candidate might be a condensate of the mobile dipoles, which carry the center of mass quantum number.  Indeed, we found that the hopping elements are proportional to $D(r)$, the dipole propagator.  If the dipoles condensed, such that $D(r)$ approached a constant as $r$ approached infinity, the fractons would appear to have a finite mass.  The problem with this scenario is that dipole condensation destabilizes the phase.  As shown in previous work\cite{genem}, a dipole $p^i$ couples to the gauge field via minimal coupling to the effective gauge field $p^jA_{ij}$.  Condensing dipoles will gap all components of the gauge field, destroy long-range entanglement, and take us to a trivial phase.

Another thought is to go to finite temperature, where we should have a thermal bath of dipoles which can be absorbed to allow fracton mobility.  Indeed, at temperatures comparable to the mass scale of dipoles, thermally excited dipoles can play a role in fracton mobility, although the physics is not quite that of free particles.  We leave the discussion of such finite-temperature physics to an upcoming work.

But there is a simpler mechanism for fracton mobility which does not require a thermal ensemble or a dipole condensate.  A fracton will be endowed with finite inertia simply by the presence of a finite density distribution of other fractons.  If space is filled with some finite density $\rho(r)$ of fractons, then a given fracton will always have particles around with which it can exchange center of mass information.  The hopping matrix elements can be calculated by summing the amplitudes for absorbing dipoles emitted from each point in space:
\begin{equation}
t(r) = \alpha \int dr' \rho(r')D(r-r')
\label{mach1}
\end{equation}
In the case where $\rho(r)$ is approximately constant, we can pull it out of the integral to obtain:
\begin{equation}
t = \alpha\rho\int dr'\,D(r-r') = \alpha\rho
\label{mach2}
\end{equation}
where we have taken advantage of the normalization and translation invariance of the propagator to show that the hopping element $t$ is now independent of $r$.  The fracton now has a finite and constant effective mass at any point in the universe, which is more in line with the normal behavior of gravitational particles.

The most notable feature of the above formulas is the fact that they provide a concrete realization of Mach's principle, a quasi-philosophical concept which historically has been difficult to put on firm footing.\cite{mach,mach2}  Loosely speaking, this principle states that inertia is not an intrinsic property of a body, but rather is given to the body by its interaction with the rest of the universe (often poetically phrased as ``the distant stars").  The idea is an appealing one, since it makes the motion of a particle inherently relative, defined with respect to other particles instead of some absolute space.  Indeed, the concept was influential to Einstein in the original development of general relativity.  The standard formulation of Einstein gravity in some sense obeys this principle, in that the large-scale distribution of matter of the universe determines its curvature, which then determines the geodesics which matter must follow.  But a precise statement of the role of Mach's principle in general relativity has been difficult to formulate.  The present treatment of gravity from the fractonic viewpoint provides clues as to how we might reformulate our understanding of gravity in a way which is more explicitly Machian, where inertial mass is not an intrinsic constant, but determined by interactions with the rest of the universe.

Considering the toy nature of the model described here, we will refrain from too much speculation about real-world gravity and cosmology, but a few basic remarks are in order.  Let us assume that our formulation of Mach's principle applies to real-world gravity.  At the current point in cosmological history, the effects would not be particularly noticeable.  Any gravitational source can generate mass for particles, and cosmological data suggests that the primary source of energy density in the present era is dark energy, which remains constant over time.  Thus, the present-day inertia of particles would be set by the dark energy content of the universe, and there would be no observed change in the inertia of particles over time.  However, things could be much different in the matter or radiation dominated eras of cosmology, when the energy density of the universe evolved as  a function of time.  At earlier times, the universe was much denser, which leads to particles having less inertial mass.  Whether or not this has any significant cosmological implications is unclear at the present time.

\subsection{The Two-Body Problem}

In order to see the gravitational attraction explicitly, we will now analyze a semi-classical limit of the two-body problem for a fracton model obeying such a ``center of mass" conservation law.  (We will not need to rely on the specifics of the toy model.)  For this type of fracton model, the gravity will generically be short-ranged, but we will discuss how the power-law behavior of Newtonian gravity can arise under certain conditions.

\subsubsection{The Basics}

We analyze the problem of two fractons of equal charge, moving via the exchange of virtual dipoles, as seen earlier in Figure \ref{fig:2part}.  These two fractons will orbit each other around their mutual center of mass.  The hopping matrix elements are non-uniform, taking the form $t(r) = \alpha D(r)$ where $D(r)$ is the dipole propagator, $r$ is the separation of the fractons, and $\alpha$ is a constant.  We note that it is not necessary to keep track of the positions of the two particles separately.  Since center of mass is always conserved, we only need to specify the position of one particle, while the other particle always stays in step on the opposite side of the center of mass.  We therefore only need to analyze a single-particle problem with hopping elements $t(r)$ depending on the distance $r$ of the particle from some fixed center of mass.

For free particles on a lattice with uniform hopping matrix elements $t$, the low-energy continuum limit of the single-particle dispersion generically takes the quadratic form:
\begin{equation}
E = -t + \frac{1}{2}tk^2a^2
\end{equation}
where $a$ is the lattice scale and $k$ is the momentum.  (The factor of $1/2$ is non-universal and is simply chosen for convenience.)  The hopping element $t$ represents the inverse effective mass.  We will now make a semi-classical approximation, where we let the effective mass depend on position, and let the momentum vary as an independent classical variable.  This is spiritually equivalent to a WKB approximation, where we allow the momentum $k$ to vary with position.  This is a rough approximation, but we will see that it is enough to get a sensible physical picture and recover the expected classical behavior of gravity.  Within this approximation (and setting $a=1$) the single-particle Hamiltonian takes the form:
\begin{equation}
H = -t(r)+\frac{1}{2}t(r)k^2 + V_0(r)
\label{disperse}
\end{equation}
where we have allowed for the possibility of a potential energy between the particles, $V_0(r)$, which is usually present in gapless fracton models as a consequence of the gapless gauge field.  One should also generically allow for tensor structure in the kinetic term, $\frac{1}{2}t_{ij}k^ik^j$, to allow for different hopping amplitudes in the radial and tangential directions.  We ignore this tensor structure for now, in order to focus on scaling, but we will comment on the tensor structure later.  We can rewrite this Hamiltonian as:
\begin{equation}
H = \frac{k^2}{2m(r)} + V(r)
\end{equation}
where the effective mass is $m(r) = 1/t(r)$ and the effective potential is $V(r) = V_0(r) - t(r)$.  The corresponding Lagrangian is:
\begin{equation}
L = \frac{1}{2}m(r)|\dot{\vec{r}}|^2 - V(r)
\end{equation}
and the equation of motion is:
\begin{equation}
\partial_t(m\dot{\vec{r}}) = -\vec{\nabla}V+\frac{1}{2}(\vec{\nabla}m)\dot{r}^2
\end{equation}
\begin{equation}
\dot{\vec{v}} = \frac{1}{m(r)} \bigg(-\vec{\nabla}V+\frac{1}{2}(\vec{\nabla}m)v^2 - (\vec{v}\cdot\vec{\nabla}m)\vec{v}\bigg)
\end{equation}
where $\vec{v}=\dot{\vec{r}}$ is the velocity.  We see that the interaction between the two fractons has three effects on the motion, all of which will contribute to an effective attraction.  First, the effective potential $V$ picks up an explicit attractive contribution, $-t(r)$.  Second, the increase in mass with separation results in an overall slowdown of the velocity as the fractons move apart, which is the same effect as an attractive interaction.  Finally, the position-dependent mass results in extra velocity-dependent forces, which we will see further amplify the increase in effective mass as the separation increases.

Finding generic solutions to this equation would be a rather daunting task.  Instead, we will content ourselves simply with understanding the behavior of circular orbits.  In this case, the $\vec{v}\cdot\vec{\nabla}m$ term vanishes (since $\vec{\nabla}m$ is radial), and the left-hand side of the equation becomes $-(v^2/r)\hat{r}$.  The radial equation of motion then becomes:
\begin{equation}
-\frac{v^2}{r} = \frac{1}{m}\bigg(-\partial_r V + \frac{1}{2}(\partial_r m)v^2\bigg)
\end{equation}
Solving for the orbital velocity, we obtain:
\begin{equation}
v = \sqrt{\frac{r\partial_r V}{m(1+\frac{r}{2m}\partial_r m)}}
\end{equation}
As promised, the mass gradient terms contribute an extra enhancement to the effective mass, by a factor of $(1+\frac{r}{2m}\partial_r m)>1$, which in the situations of interest will increase with separation.

Let us first consider the case where there is no intrinsic interparticle potential, $V_0(r) = 0$, as would be the case in a gapped fracton model.  This will allow us to focus on the physics which is intrinsic to fractons, without worrying about the specifics of the gauge field.  In this case, the potential reduces to $V(r) = -t(r)$, so the orbital velocity becomes:
\begin{equation}
v = \sqrt{\frac{-rt\partial_r t}{1-\frac{r}{2t}\partial_r t}}
\label{orbv}
\end{equation}
All that remains now is to plug in the appropriate hopping matrix element $t$.  We will begin by assuming that the dipoles mediating the motion have some energy gap $M$, which is the generic situation in fracton models.\cite{foot8}  In this case, we have exponential decay of the matrix elements, $t(r)= be^{-Mr}$ (for some constant $b$).  Using this value in our orbital velocity, we obtain:
\begin{equation}
v(r) = be^{-Mr}\sqrt{\frac{Mr}{1+\frac{1}{2}Mr}}
\end{equation}
The orbital speed decays to zero exponentially as the fractons separate.  When the fractons are separated by more than a few times $M^{-1}$, motion is completely negligible, as we expect for fractons.

We can also consider the case where there is a finite background density, such that we have a nonzero asymptotic hopping matrix element, $t(r) = t_0 + b e^{-Mr}$, in line with our discussion of Mach's principle.  In this situation, fractons can behave as free particles at infinity.  Nevertheless, the fractons will still attract, and we can consider the properties of bound states.  The orbital velocity in this case will take the long-distance form:
\begin{equation}
v \approx e^{-Mr/2} \sqrt{t_0bMr}
\end{equation}
which once again decays exponentially as the fractons separate.  In this case, however, the asymptotic inertia of fractons allows them to exist in unbound states of only slightly higher energy.  The above velocity profile corresponds to weakly bound states, which can be unbound via a fairly small kick to one of the fractons.

\subsubsection{Towards Newtonian Gravity}

The exponential dropoff of motion seen above seems in line with the expected behavior of fractons, but it is certainly not in line with the expected behavior of gravity in the real world, where the gravitational force is a power law.  This leads us to consider an alternative case, where the dipoles mediating the motion are gapless.  In the presence of gapless dipoles, the hopping matrix elements $t(r)$ will behave as a power law, giving rise to a power-law gravitational attraction.  However, the precise calculation of $t(r)$ is more subtle in this case, and we will not be able to carry out the analysis exactly.  Nevertheless, we will see that certain reasonable assumptions can give us the correct scaling of Newtonian gravity.  Ultimately, however, the results of the present section need to be backed up by a more in-depth investigation of gapless dipole models in the future.

\begin{figure}[b!]
 \centering
 \includegraphics[scale=0.35]{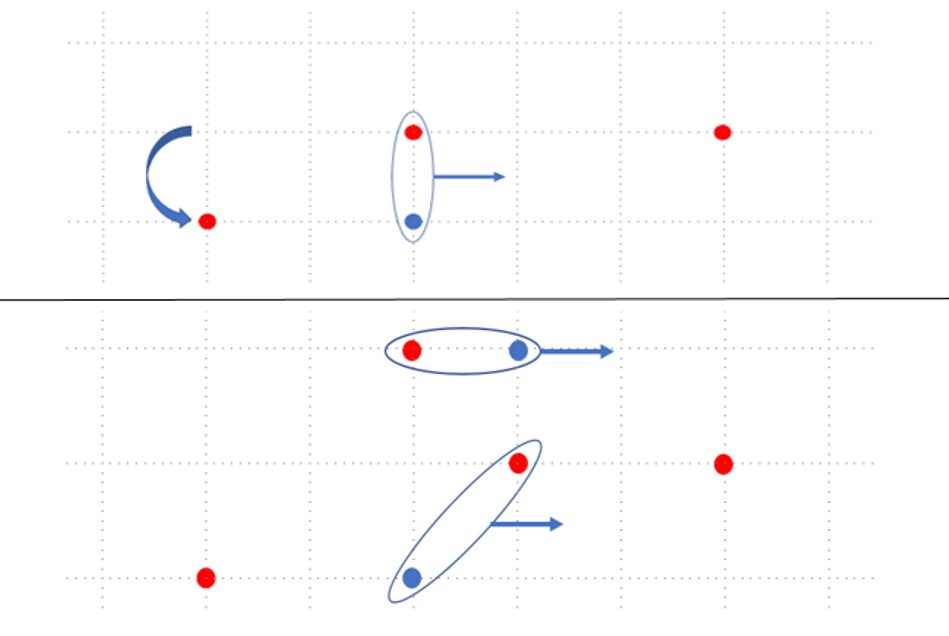}
 \caption{A fracton hops by emitting a dipole which begins to propagate.  The emitted dipole can change its moment by emitting a second dipole.  When the dipoles are gapless, such branching processes have non-negligible amplitude.}
 \label{fig:split}
 \end{figure}

In the presence of gapless dipoles, fracton motion still occurs via the exchange of virtual dipoles.  However, there is a more complicated set of intermediate exchange processes to deal with.  For example, a virtual dipole can branch into two different species of dipoles before absorption by the second fracton, as seen in Figure \ref{fig:split}.  Even more complicated branched multi-body processes are possible.  When the dipoles are gapless, these processes are comparable in weight to single-dipole propagation and must be taken into account.  Furthermore, all of this propagation is taking place in a background power-law electric field $E_{ij}$ from the two fractons, further adding to the complexity of the problem.

It is unclear how to directly calculate $t(r)$ in this situation, but we can make a heuristic argument for the scaling.  The branched pathways that the dipoles follow also happen to correspond to the generalized ``field lines" of the electric tensor $E_{ij}$, $i.e.$ continuous configurations obeying $\partial_i\partial_j E^{ij} = 0$ away from the charges.  To get the hopping element $t(r)$, we essentially need to count the electric field lines starting at the location of one fracton and passing through the location of the other, which is equivalent to finding the average electric tensor felt by a particle due to the other.  We therefore should have:
\begin{equation}
t(r)\propto |E_{ij}(r)|\propto \frac{1}{r}
\end{equation}
using the fact that the electric tensor for a point charge scales as $1/r$ in this model.\cite{sub,genem}  We conjecture that the more precise relationship $t_{ij}\propto E_{ij}$ holds.  This is, of course, a very heuristic derivation, which should be backed up by more concrete calculations.  Nevertheless, we will see that this behavior is precisely what is needed both to recover Newtonian gravity and to connect with the geodesic principle, as we will see shortly.

We have argued that, when dipoles are gapless, hopping matrix elements behave as $t(r) = b/r$ for some constant $b$.  Let us work directly with the generic case of a finite background density, so that $t(r) = t_0 + b/r$.  Plugging this into Equation \ref{orbv} for the orbital velocity, we obtain the long-distance behavior:
\begin{equation}
v \approx \sqrt{\frac{bt_0}{r}}
\end{equation}
which scales precisely like the Keplerian velocity profile expected from the inverse square law of Newtonian gravity.  Recovering this behavior is comforting and gives us some confidence in our conjectured scaling of $t(r)$.  Note that, if $t_0$ were not present, we would actually end up with $1/r$ behavior for the velocity.  The presence of the finite density background is actually crucial for correctly recovering Newtonian gravity.

The above analysis has indicated that obtaining Newtonian gravity requires gapless dipoles, whereas the gapped case leads only to short-ranged attraction.  How are we to make sense of this?  Since power-law gravity seems to require gapless dipoles, is there some mechanism which could enforce this behavior in a real gravity model, without relying on fine-tuning?  One might consider some form of symmetry protection which could keep the dipole dispersion gapless, but this is a rather unattractive feature.  It would seem odd if the stability of Newtonian gravity relied on a special symmetry of some underlying lattice.  Luckily, the full nonlinear version of Einstein gravity nicely circumvents this problem.  The nonlinearity implies that the graviton itself acts as a source for the gravitational field.  In other words, the graviton carries the center of mass quantum number.  The gaplessness of the graviton is protected by the gauge symmetry of the theory, providing a natural source of gapless dipoles which can be exchanged between particles.  This feature is absent in the linearized gravity theory considered here, but in a more complete theory of quantum gravity, nonlinearity and gauge invariance can combine to guarantee the existence of gapless dipoles which can mediate an effective inverse square force.

\subsubsection{Fractons with Gauge Potential}

So far, we have only considered the intrinsic attraction between fractons arising from the center of mass conservation law.  For gapped fracton models, this story is fairly complete, and we can unambiguously say that such fractons will attract each other.  But, as mentioned earlier, for gapless fracton models one must also typically include an ``electromagnetic" potential arising from the gapless gauge field, which may or may not modify the story in certain regimes, depending on the form of the potential.  For example, in the toy model discussed earlier, the gauge potential between two identical fractons is repulsive, taking the form $V(r) = -\lambda r$ for constant $\lambda$.\cite{genem,foot9}  At short distances, the linear term is negligible, and we maintain an attractive interaction.  At long distances, however, the repulsive linear term is dominant, destabilizing any circular orbits.

This long-distance repulsion is not a generic feature of fracton models.  The gauge potential between like charges is repulsive on very general grounds.  However, in some other fracton models\cite{gu,gu2}, the potential decays as $V(r)\propto 1/r$.  (Indeed, the same type of behavior is seen in the Hamiltonian formulation of linearized Einstein gravity.\cite{gr})  As long as the coefficient is not too large, adding such a repulsive $1/r$ term does not result in long-distance repulsion of fractons, but simply modifies orbital velocities.  The long-distance repulsion from a linear potential is an oddity of the toy model which is not shared by more general gravitational models.

\subsection{On the Geodesic Principle and the Emergence of Space}

In studying the two-body problem, we argued that the effective Hamiltonian for a fracton takes the form:
\begin{equation}
H_{frac} = \frac{1}{2}t_{ij}k^ik^j + V
\end{equation}
where we have restored the tensor nature of the hopping.  Going to the Lagrangian formalism, the appropriate action is:
\begin{equation}
S_{frac} = \int dt \bigg(\frac{1}{2}t^{-1}_{ij} \dot{x}^i\dot{x}^j - V\bigg)
\label{fracact}
\end{equation}
where $x$ denotes the particle's coordinates.  This should be compared with the standard geodesic principle of general relativity.  The geodesic equation is most easily obtained by applying the variational principle to the proper length of a particle's trajectory\cite{foot10}:
\begin{equation}
S_{geo} = \int \sqrt{-g_{\mu\nu}dx^\mu dx^\nu}
\label{geode}
\end{equation}
Let us linearize around a flat background and take a quasistatic metric, such that $g_{0i}=0$.  The relevant degrees of freedom are then the metric perturbations $h_{ij}$ and $h_{00}$.  In terms of these variables, and taking the non-relativistic limit, the action behaves as:
\begin{equation}
S_{geo} \approx \int dt \bigg((a\delta_{ij} + h_{ij})\dot{x}^i\dot{x}^j - h_{00}+\cdot\cdot\cdot \bigg)
\end{equation}
where we have kept a scale factor $a$ setting the size of the spatial part of the flat metric.  We can now see that the geodesic action matches up fairly well with the fractonic action, Equation \ref{fracact}, provided we make two identifications.  First, we must identify $h_{00}$ as giving the potential energy, which is a familiar correspondence from the non-relativistic limit of Einstein gravity.  Next, we must also identify the inverse hopping elements as our effective spatial metric:
\begin{equation}
\frac{1}{2}t_{ij}^{-1} = g_{ij} = a\delta_{ij} + h_{ij}
\end{equation}
We argued earlier that, in the presence of a background density $\rho$, the hopping elements near a point charge behave as:
\begin{equation}
t_{ij}(r) = \alpha\rho \delta_{ij} + \hat{t}_{ij}(r)
\end{equation}
where $\hat{t}_{ij}$ is the contribution from the point charge, decaying to zero as $r\rightarrow\infty$, which serves as a perturbation to the background hopping element.  At large $r$, we can Taylor expand the inverse to obtain:
\begin{equation}
t^{-1}_{ij}(r) = \frac{1}{\alpha\rho}\delta_{ij} - \frac{\hat{t}_{ij}}{(\alpha\rho)^2}
\end{equation}
This is to be compared with the metric $(a\delta_{ij}+h_{ij})$.  The second term is fairly straightforward to interpret as the metric perturbation.  Indeed, in the case of gapless dipoles, we argued that $\hat{t}_{ij}$ was proportional to $E_{ij}$ of the point charge, decaying as $1/r$, which is the same as the long-distance behavior of $h_{ij}$ due to a point charge in Einstein gravity.  (Recall that $E_{ij}$ has served as a proxy for $h_{ij}$ in our toy model.)  We therefore see that, barring one final subtlety, the action for fractons will lead to a ``geodesic" principle of the same structure as in Einstein gravity.\cite{foot11}

This final subtlety lies in the flat part of the effective metric.  While the perturbation term was straightforward to interpret, the background term is a bit more interesting.  The presence of a flat background in the metric relied on having a finite density of fractons.  The geodesic formalism suggests that the flat part of the effective metric should behave as $\frac{1}{\rho}\delta_{ij}$, so that the matter content sets the size of the emergent universe relative to the lattice scale.  When there is no finite density of particles, the metric blows up and all points are infinitely far away from each other within the emergent universe.\cite{foot12}  This is a complementary viewpoint to our earlier discussion in terms of inertia.  Instead of viewing an isolated particle as infinitely massive, we can view it as being infinitely far away from all other points in space, leading to the same immobility.

While equivalent to our previous discussion, the present perspective gives us an interesting way to understand the emergence of space.  Without a finite density of particles, we no longer possess a smoothly connected space, but rather a set of independent isolated points, infinitely far away from each other.  It is only when we excite the system, filling the emergent universe with particles, that these points coalesce into a connected space and we recover a useful sense of geometry, via the inverse metric picking up an expectation value.  As in our discussion of Mach's principle, it is not clear to what extent this is an entirely ``new" idea in a gravitational context.  The principle of matter determining the structure of space is already contained in Einstein gravity.  But the formalism described here gives us a more explicit understanding that matter is constructing the space around it, as opposed to simply perturbing a pre-existing background.

In this light, a flat background space should not be a fundamental object which we need to put into a gravitational theory by hand, but rather is an emergent concept which is collectively produced by a large number of particles, mutually generating a notion of proximity with each other.  This is essentially Mach's principle taken to the extreme: the ``distant stars" not only give inertia to particles, but also create the space in which those particles can move.  This gives yet another reinterpretation of the fracton phenomenon.  Fractons cannot move because they do not possess a connected background space to move through.

Of course, it would be nice to have a more concrete mathematical formulation of the principle of matter creating space, without reference to the pre-existing geometry of the lattice of our system.  How to formulate such a theory of emergent geometry without relying on either a flat background or an embedding space is not entirely clear.  One would need to start from a set of disconnected points and construct space from scratch, perhaps through some information-theoretic entanglement measure, such as in Reference \onlinecite{carroll}.  Formulating a theory without the crutch of a background space will likely prove to be a formidable challenge.

\section{Conclusion}

In this work, we have laid the groundwork for a conceptual reconciliation of fractons and gravity.  We examined a toy model consisting of fractons coupled to an emergent graviton, and we resolved how the fracton phenomenon is consistent with the expected behavior of gravity.  We showed that the conservation laws leading to fractonic behavior have a natural place in a gravitational theory in the form of conservation of center of mass.  Fracton mobility arises via the virtual exchange of propagating particles carrying the center of mass quantum number.  Center of mass can be exchanged more efficiently when fractons are close together, leading to an effective gravitational attraction between them.  We argued that gravitational attraction is a natural feature of models with local center of mass conservation.  We also showed how fractons can be endowed with a finite inertia in the presence of a finite density distribution of other fractons, in a direct manifestation of Mach's principle.  We analyzed orbital motion, formulating an appropriate geodesic principle and discussing how to obtain power-law Newtonian gravity.  We have also put forward ideas on how space itself is created by the particles existing within it.

But there remains much that could be done.  The model we focused on here only had an emergent energy quantum number.  For a more realistic gravity theory, we should also include an emergent momentum quantum number.  We should also work to formulate a fully nonlinear theory, where the graviton serves as a gravitational source.  On the technical side, some of the calculations done here require a more systematic framework, such as the calculation of hopping matrix elements in the presence of gapless dipoles.  Besides such technical work, there also remains a whole host of interesting gravitational problems which might be productively studied from the present fractonic viewpoint.  Can we obtain black holes, for example?  Such questions remain as challenges for the future.

\section*{Acknowledgments}

Thanks are due to Shoucheng Zhang, Brian Swingle, and Yahui Zhang, who provided stimulating conversations which heavily influenced the present work.  Particular thanks are due to Brian Swingle for providing valuable feedback on an early version of this paper.  I also would like to thank Xiaoliang Qi, Liujun Zou, Sagar Vijay, T. Senthil, Debanjan Chowdhury, Inti Sodemann, Samuel Lederer, Rahul Nandkishore, Mike Hermele, Han Ma, and Abhinav Prem for useful discussions.  I acknowledge monetary support from the Simons Foundation through a Simons Investigator award to T. Senthil.  I also acknowledge TA funding from the Department of Physics at MIT.  On that note, I would like to thank all of my 8.02 students for a very enjoyable semester.

\section*{Appendix A:  The Lattice Model}

The model we have considered in this paper features a symmetric tensor gauge field $A_{ij}$.  Putting such a tensor on a lattice is not quite as simple as for a vector gauge field, where the field quite naturally lives on links.  The simplest choice of lattice model was described in References \onlinecite{alex,cenke1,cenke2}, with some additional details discussed in References \onlinecite{sub,genem}.  In these models, we work with a cubic lattice, on which we will put six independent components of the symmetric tensor: $A_{xx}$, $A_{yy}$, $A_{zz}$, $A_{xy}$, $A_{yz}$, and $A_{xz}$.  Since we generally allow the gauge field to be compact, each of these components is a quantum rotor: a $U(1)$ variable, taking values between $0$ and $2\pi$.  This is in contrast to a noncompact gauge field, where the components would simply be real-valued.  One important consequence of compactness is that the canonical conjugate variables to $A_{ij}$ must be quantized to integer values, since they are essentially the angular momenta of the rotors.  We call these conjugate variables $E_{xx}$, $E_{yy}$, $E_{zz}$, $E_{xy}$, $E_{yz}$, and $E_{xz}$.

\begin{figure}[b!]
 \centering
 \includegraphics[scale=0.3]{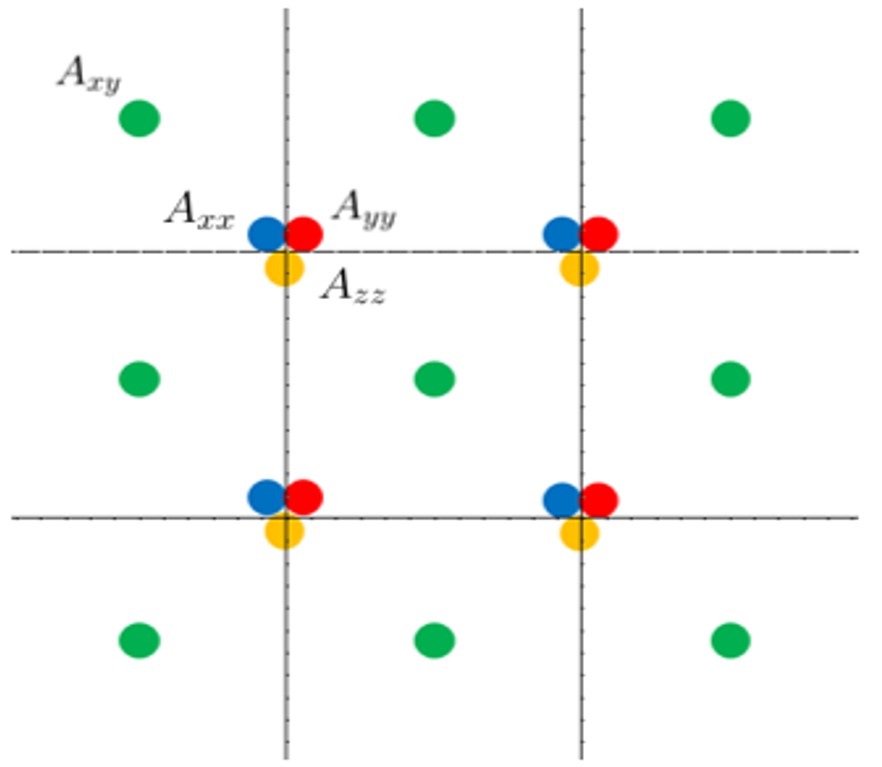}
 \caption{We here illustrate a planar cross-section (in the $xy$ plane) of the cubic lattice on which our model is defined.  The off-diagonal $A_{xy}$ rotor lives on each of the plaquettes.  The remaining off-diagonals live in other planar cross-sections, not pictured here.  All diagonal elements ($A_{xx}$, $A_{yy}$, $A_{zz}$) live on each vertex.  (We have spread out the three colored circles on each vertex simply for ease of illustration.  All three rotors exist at the same location.)}
 \label{fig:lattice}
 \end{figure}
 
To gain some intuition regarding where the components of the tensor should live, let us examine the behavior of the simplest possible rank 2 tensor: a second derivative.  Consider a function $\alpha(r)$ defined on the vertices of the cubic lattice.  The first derivatives, $\partial_i\alpha$, will all live on links of the cubic lattice, with the $x$ component living on links in the $x$ direction, and so on.  A diagonal second derivative, such as $\partial_x\partial_x \alpha$, represents the difference of two links lined up end to end, and will therefore naturally live on a vertex.  An off-diagonal derivative, such as $\partial_x\partial_y\alpha$, represents the difference of two links lined up side by side, and will naturally live on the faces (plaquettes) of the lattice.  Using this intuition, we put the diagonal components of our gauge field ($A_{xx}$, $A_{yy}$, $A_{zz}$) on the vertices of the lattice, and we put off-diagonal elements on the appropriate faces.  For example, $A_{xy}$ lives on plaquettes in the $xy$ plane, and so on.  This behavior is illustrated in Figure \ref{fig:lattice}.  Note that there are three rotors on each vertex.  Nevertheless, the three rotors are not identical at the level of Hamiltonian, as the $A_{xx}$ component will have interactions predominantly in the $x$ direction, and so on.  The situation is quite similar to $p$ orbitals on atoms in a cubic lattice environment.  Indeed, we expect $p$ orbitals to be useful in the search for solid-state realizations of this phase.

As mentioned in the main text, the primary physics of the phase is determined by the Gauss's law, which is an energetically imposed constraint on the system.  For the phase in question, the desired ground state constraint is:
\begin{align}
\begin{split}
\partial_i\partial_j E^{ij} =& \\
\partial_x\partial_xE^{xx} + &\partial_y\partial_yE^{yy} + \partial_z\partial_zE^{zz} +\\
2(\partial_x\partial_y&E^{xy} + \partial_y\partial_zE^{yz} + \partial_x\partial_zE^{xz}) = 0
\end{split}
\end{align}
We therefore include an energetic penalty term in the Hamiltonian of the form:
\begin{align}
\begin{split}
H_U = U&(\partial_x\partial_xE^{xx} + \partial_y\partial_yE^{yy} + \partial_z\partial_zE^{zz} +\\
&2(\partial_x\partial_yE^{xy} + \partial_y\partial_zE^{yz} + \partial_x\partial_zE^{xz}))^2
\end{split}
\end{align}
where the derivatives should all be understood as lattice derivatives, such as:
\begin{equation}
\partial_x\partial_x E^{xx} = \frac{1}{a^2}(E^{xx}(x-a) - 2E^{xx}(x) + E^{xx}(x+a))
\end{equation}
with $a$ being the lattice spacing.  We also include in our Hamiltonian the lowest order energy term for the electric field:
\begin{equation}
H_E = \frac{1}{2}E^{ij}E_{ij}
\end{equation}
(The factor of $\frac{1}{2}$ is purely for convenience.)  We also wish to include terms involving the field $A_{ij}$ directly.  In the low energy sector, the only terms that are relevant are those which commute with the gauge constraint.  All other terms can be projectively eliminated from the effective theory for the low-energy sector (in essentially the same way that one goes from the half-filled Hubbard model to the $tJ$ model by projectively eliminating double occupancies).  For a term in the Hamiltonian to commute with the gauge constraint, an equivalent statement is that it must be invariant under the corresponding gauge transformation.  Since $A_{ij}$ and $E_{ij}$ are canonically conjugate variables, the constraint $\partial_i\partial_j E^{ij} = 0$ implies gauge invariance under the transformation:
\begin{equation}
A_{ij} \rightarrow A_{ij} + \partial_i\partial_j \alpha
\end{equation}
where, once again, all derivatives should be interpreted as lattice derivatives.  Just as in ordinary electromagnetism, we need to find some ``magnetic field" object which is invariant under the gauge transformation.  It can be explicitly checked that the lowest order magnetic object which can be constructed takes the form\cite{foot13} $B_{ij} = \epsilon_{iab}\partial^aA^b_{\,\,\,j}$.  We then include a term in the Hamiltonian of the form:
\begin{equation}
H_B = -\sum_{ij} \cos(B_{ij}) \approx -1 +\frac{1}{2}B^{ij}B_{ij}
\end{equation}
We started with a cosine, in order to respect compactness.  In the deconfined phase of the gauge theory, we can expand the cosine around its minimum\cite{alex}, in close analogy with the standard treatment of compact $U(1)$ lattice gauge theory, to obtain a simple quadratic term.  At the end of the day, the full Hamiltonian reads (up to constants):
\begin{equation}
H = \frac{1}{2}(E^{ij}E_{ij} + B^{ij}B_{ij}) + H_U + \cdot\cdot\cdot
\end{equation}
where ``$\cdot\cdot\cdot$" represents terms that don't commute with the gauge constraint and are therefore irrelevant to the low-energy physics.  In other words, these are the ``non-gauge-invariant" terms, which many would simply regard as non-existent.  However, as emphasized elsewhere \cite{u1}, it is useful to regard gauge symmetry as only being emergent in the low-energy gauge sector, with high-energy violations of the gauge constraint being regarded as massive charges.  In that light, these terms simply represent the piece of the Hamiltonian governing the dynamics of the massive charges.

For example, consider the non-gauge-invariant operator $e^{iA_{xy}}$, which serves as a raising operator, increasing $E_{xy}$ by one.  This takes us out of the pure gauge sector, or in other words, creates particles.  In fact, this operator creates precisely the square quadrupole configuration seen in Figure \ref{fig:quadrupole1}.  Similarly, the operator $e^{iA_{xx}}$ raises $E_{xx}$ by one.  The result is the linear quadrupole configuration seen in Figure \ref{fig:quadrupole2}.  Similar behavior is found for all other components of $A_{ij}$, with the quadrupoles aligned in different directions.

When the gauge field is compact, an important question exists regarding the stability of these phases.  For example, in $(2+1)$ dimensions, a standard compact $U(1)$ gauge theory is unstable to confinement.\cite{polyakov}  This is because the magnetic portion of the action admits instantons (``monopoles"), point-like spacetime defects, which always proliferate in the path integral, gapping the photon and confining the charges, thereby destroying the phase.  (The noncompact version of the theory does not share this instability, due to the lack of monopoles.)  In the $(3+1)$-dimensional compact $U(1)$ gauge theory, on the other hand, the magnetic monopoles are solitons: points in space, lines in spacetime.  Solitons differ from instantons in that we can specify dynamics for these particles.  When the monopoles are gapped, there is no proliferation of magnetic defects, and the phase is stable.  It is only when the monopoles condense that the photon is gapped and the phase is destroyed.  For this reason, the $(3+1)$-dimensional $U(1)$ spin liquid has a stable deconfined phase, whereas the $(2+1)$-dimensional analogue is inherently unstable to confinement in the absence of gapless charges.  Similar logic holds in the present case of a symmetric tensor gauge theory.  One can explicitly check that the magnetic defects are solitons, given by $\partial_i B^{ij} = \tilde{\rho}^j$ for magnetic vector charge $\tilde{\rho}^j$.\cite{genem}  When these magnetic charges are gapped, the phase is stable.  Only by condensing the magnetic charges do we gap the gauge field and destabilize the phase.  (It is unclear if these magnetic defects play any interesting role in the gravitational dynamics of this phase.  These defects are actually 2-dimensional particles, restricted to move transversely to their charge vector, so they do not have the correct structure to play the role of the emergent momentum.  We have ignored these particles throughout the present work, since their inclusion should not modify the central message of the story.  In any case, we can always let such magnetic particles have a much larger gap than electric charges.)

This phase also has a set of generalized Maxwell equations, which amount to our toy version of Einstein's equations.  We will not go into the derivation here, referring the reader to Reference \onlinecite{genem} for details.  We simply state the result:
\begin{equation}
\begin{split}
\partial_i\partial_j E^{ij} &= \rho\\
\partial_i B^{ij} &= \tilde{\rho}^j \\
\epsilon^{iab}\partial_a E_b^{\,\,\,j} &= \partial_t B^{ij} + \tilde{J}^{ij}\\
\frac{1}{2}(\epsilon^{iab}\partial_a B_b^{\,\,\,j} + \epsilon^{jab}\partial_a B_b^{\,\,\,i}) &= -\partial_t E^{ij} - J^{ij}
\end{split}
\end{equation}
The top two equations represent the electric and magnetic Gauss's laws, and the bottom two represent the generalized Ampere and Faraday laws.  Note that particles in this theory no longer have a vector current $J^i$ representing the motion of individual particles.  Rather, there is a tensor current $J^{ij}$ representing two-body hopping processes, which serves as the analogue of the stress tensor $T^{ij}$ in this model.  The first and fourth equations above (involving $\rho$ and $J^{ij}$) are the direct analogue of Einstein field equations, namely the $00$ and $ij$ components.  There is no analogue of the $0i$ components due to the lack of an emergent momentum quantum number.  The symbol $\tilde{\rho}^j$ denotes the magnetic charge, and $\tilde{J}^{ij}$ is the corresponding magnetic current.  These are only nonzero when the gauge field is compact.  In the noncompact limit, these terms vanish and the corresponding Maxwell equations become trivial, which is why these extra equations are usually not included as Einstein equations.

\section*{Appendix B:  The Weinberg-Witten Theorem}

A standard line of thinking holds that we cannot obtain emergent gravity within the same spacetime as some original nongravitational degrees of freedom.  This is largely due to the Weinberg-Witten theorem, which roughly says (among other things) that a Lorentz-invariant quantum field theory with a conserved stress-energy tensor $T^{\mu\nu}$ cannot have massless excitations carrying spin higher than 1.\cite{ww}  At first glance, this seems to rule out the possibility of gravity emerging within a given background spacetime.  Thus, much more focus has been placed in recent years on holographic theories, where the emergent gravity resides in a dual asymptotically AdS spacetime.  The holographic framework is quite useful, giving us valuable insight into the structure of quantum gravity, but many principles do not carry over directly to flat or de Sitter spacetime.\cite{foot14}  In this work, however, we have worked with a model where gravitational structure explicitly emerges as the low-energy theory of the original flat $(3+1)$-dimensional spacetime.  This surely causes some discomfort to many.  Shouldn't the Weinberg-Witten theorem have prevented this type of model from existing?  The answer is that fracton models are outside the jurisdiction of the Weinberg-Witten theorem, by violating some of its fundamental assumptions, and the limitations of this theorem do not restrain us here.

To begin with, there are some obvious (though not particularly strong) reasons why the toy model discussed here is outside the domain of Weinberg-Witten.  One could cite the fact that the toy model is not fully Lorentz-invariant.  One could also cite the fact that, in the linearized model considered here, the graviton did not carry the gravitational charge.  Technically speaking, these details already free us from the Weinberg-Witten theorem.  However, as we have discussed, appropriate modifications should allow both emergent Lorentz invariance and nonlinearity to appear in the theory, so we should not put much weight on these lines of argument.

As physicists, it is much more important for us to understand why the \emph{spirit} of Weinberg-Witten is being violated.  At the end of the day, our circumvention of the Weinberg-Witten theorem is permitted by the nature of fractons.  The proof of the Weinberg-Witten theorem relies heavily on the use of asymptotic single-particle momentum eigenstates.  In a fracton model, however, a charged particle with finite momentum can only exist when the particle is in the middle of a finite density bath of other charges.  (Recall that ``charge" refers to energy in the gravitational context.)  In the asymptotic regime, where all particles are well-separated, charges lose their mobility and momentum states no longer exist as well-defined excitations of the system.  The lesson we learn from the fractonic viewpoint is that momentum eigenstates are not fundamental asymptotic single-particle excitations of a gravitational system, but rather only exist as an effective description of particles in a finite-density system.

Similarly, for gauge theories of spin higher than 2, where the conservation laws are even more restrictive, it is questionable whether there exists anything even resembling a momentum eigenstate.  For these theories, the fundamental formulation really occurs in position space.  The insights gained from fractons therefore change our understanding of the Weinberg-Witten theorem.  It is not the case that interacting massless higher spin theories don't exist, but rather that these theories do not admit well-defined asymptotic momentum eigenstates for their charges, which is a special feature of lower-spin gauge theories.


\begin{thebibliography}{9}
\bibitem{foot1}
The term ``fracton" was used in earlier statistical mechanics literature to mean something completely different.
\bibitem{foot2}
There are other gapless gauge modes in addition to the helicity $\pm 2$ modes.  These could be eliminated by more complicated gauge constraints, if so desired, but their presence will not affect any of our arguments, so we stick with the simplest theory.
\bibitem{foot3}
We have drawn these operators for a cubic lattice model, as discussed in Appendix A, but the lattice details will not be important for the conclusions here.
\bibitem{foot4}
There are subtle differences between a pure transformation on the metric and a true diffeomorphism, which also transforms coordinates, but these differences will not be important for the conclusions made here.
\bibitem{foot5}
The literature on the quantization of linearized gravity is too vast to be cited both accurately and usefully.  For the sake of clarity, we refer the reader to Reference \cite{gr}, which was deemed to be the most useful.  Additional references can be found therein.
\bibitem{foot6}
Technically, it is ``center of energy," but this reduces to center of mass at low energies, and we will stick with this more standard terminology.
\bibitem{foot7}
Technically, this should be the initial distance between them, as opposed to the final distance, but this distinction should be unimportant as long as the particle separation is much larger than the lattice spacing.
\bibitem{foot8}
One must resist the temptation to use ``gap" and ``mass" synonymously.  The inertial mass can be very different from the energy gap in the present context.
\bibitem{foot9}
The presence of such a linear potential is surely distressing to some, since linear potentials normally indicate confinement.  We refer the reader to previous works \cite{sub,genem} for a more detailed discussion of this ``electrostatic confinement" issue, where we find that the fractons are actually stable excitations, as the rules for confinement are slightly different in fracton models.
\bibitem{foot10}
One usually takes the action to also have a factor of particle mass $m$, which in the standard treatment is a constant.  But within the framework discussed in this paper, particles have no intrinsic inertial mass $m$, but rather an inertia determined by the background around it.  In keeping with this philosophy, we consider the factor of mass to be part of the dynamical metric, so that the effective metric $g_{\mu\nu}$ has dimensions of mass squared.
\bibitem{foot11}
Of course, the formulation here is not directly in terms of a proper length, so the term ``geodesic" is not entirely appropriate.  Still, the particle moves along ``extremal" curves determined by the metric.
\bibitem{foot12}
The statement that each point is infinitely far from its nearest neighbor relies on having some underlying scale, either atomic or Planckian, on which space is discrete, so that the concept of nearest neighbor exists.  This is manifestly true in a condensed matter context, and there are reasons for thinking this may hold for real-world gravity as well.
\bibitem{foot13}
Note that, unlike all other variables used so far, this is a non-symmetric tensor.  This is related to the fact that this theory is not fully self-dual.  We refer the reader to Reference \onlinecite{genem} for further discussion.
\bibitem{foot14}
For example, the number of bulk degrees of freedom in AdS space is of the same order as the number of degrees of freedom on the boundary, which enables a duality.  In flat or de Sitter space, on the other hand, the bulk has far more degrees of freedom than the boundary, so there is no real reason to suspect such dualities to work.
\end{thebibliography}
\end{document}